 \documentstyle{article}
 \include{matex}
\begin{document}
 \begin{titlepage}
 \begin{center}
 \rightline{FTUV/92-20}
 \rightline{IFIC/92-21}
 \rightline{LMU-05/92}
 \rightline{June 1992}
 {\Large \bf OBSERVABLE MAJORON EMISSION IN NEUTRINOLESS DOUBLE BETA DECAY}
 \vs .2cm
 {\large \bf Z. G. Berezhiani}$^{1}$ \\
 \vs .2cm
 {\it Sektion Physik, Universitat Muenchen, 8000 Muenchen 2, Germany and \\
 Institute of Physics of the Georgian Academy of Sciences,
 SU-380077 Tbilisi, Georgia\\}
 \vs .2cm
 {\large \bf A. Yu. Smirnov}$^{2}$\\
 \vs .2cm
 {\it Institute for Nuclear Research, Russian Academy of Sciences,
 SU-117312 Moscow, Russia}\\
 and \\
 \vs .2cm
 {\large \bf J. W. F. Valle}$^3$ \\
 \vs .2cm
 {\it Instituto de F\'{\i}sica Corpuscular - C.S.I.C.\\
 Departament de F\'isica Te\`orica, Universitat de Val\`encia\\
 46100 Burjassot, Val\`encia, SPAIN}\\
 \end{center}

 \begin{quotation}
 We consider a class of simplest Majoron models where
 neutrino-majoron couplings can be in the range
 $g \sim 10^{-5} - 10^{-3}$, leading to the observability
 of neutrinoless double beta decay with majoron emission.
 The majoron is a singlet (or mostly a singlet) of the
 electroweak \21 gauge symmetry, thus avoiding conflict
 with the LEP data on Z-decay, which rules out the triplet
 and doublet majoron models. Some aspects of the phenomenology
 of such models are discussed.
 \end{quotation}
 \vfill
 $^1$ {\it Alexander von Humboldt Fellow. Internet
 ZURAB@hep.physik.uni-muenchen.de \\}
 $^2$ {\it Internet SMIRNOV@inucres.msk.su \\}
 $^3$ {\it Bitnet VALLE@EVALUN11 - Decnet 16444::VALLE\\}
 \end{titlepage}

 \section{Introduction}

 An attractive approach to the neutrino mass problem is to consider
 neutrinos as Majorana particles with small masses arising as a
 result of spontaneous lepton number violation \cite{fae}. In these
 models a massless Goldstone boson, the majoron should exist.
 Several realizations of this idea were proposed. They are
 distinguished mainly by the weak isospin (I) properties of the
 majoron and have different phenomenology: the singlet majoron
 with I=0 \cite{CMP}, the triplet one with I=1 \cite{GR} and
 the doublet majoron I=1/2 \cite{ARCA}.

 If the majoron Yukawa couplings with neutrinos are sufficiently strong
 $(g \sim 10^{-5} - 10^{-3})$, this would imply several interesting
 consequences for particle physics and astrophysics. Perhaps the most
 striking is the possibility of substantial neutrinoless double beta
 decay rates with majoron emission, denoted ($2 \beta_{0 \nu J}$)
 \cite{GGN}. On the other hand, the majoron can play a significant
 role for the history of the Early Universe, in the evolution
 of stellar objects, and in supernovae astrophysics. Neutrino
 decay with majoron emission may also be relevant in connection
 with the solar neutrino problem \cite{uses2,uses1,Ra,zurab}. Finally,
 the existence of neutrino masses in excess of 50 eV or so, \sa the
 17 KeV neutrino \cite{SIMPSON,Hime91,17exp}, if confirmed,
 can be considered as a strong argument in favour of the existence
 of such a majoron, since only fast decays (or annihilation) of
 such neutrinos into majorons can avoid conflict with cosmological
 and astrophysical restrictions \cite{fae,GNP}.

 The existence of majorons with nontrivial weak isospin properties,
 \sa the triplet and doublet majorons would lead to appreciable
 neutrino-majoron ($\nu$-J) couplings. However these are strongly
 excluded by recent LEP data \cite{LEP1} on the invisible width of
 the Z-boson. Indeed, they would contribute to this width through
 the decay $Z \ra \rho + J$, where $\rho$ is a scalar partner of
 the majoron J, the equivalent of 2 and 1/2 extra neutrino species,
 respectively, whereas the LEP limit on the number of active light
 neutrino species %%$N_{\nu} = 3.04 \pm 0.05 $,
 allows only the three known neutrinos, to a few percent accuracy.
 On the other hand the singlet "seesaw" majoron \cite{CMP} is
 extremely weakly coupled with neutrinos. This would suggest that
 it is meaningless to search for the effects of a large $\nu$-J
 coupling, \sa the search for $2 \beta_{O \nu J}$ decay.

 In this letter we show that there is no direct contradiction
 between the absence majoron emission in Z-boson decay and the
 possibility of fast $2 \beta_{0 \nu J}$ decay. The former is
 determined by a gauge coupling and the weak isospin content
 of the majoron, whereas the latter is related only to the
 majoron Yukawa coupling. We present a variety of simple models
 in which the majoron, being a singlet or predominantly singlet
 of the electroweak \21 symmetry, nevertheless can be sufficiently
 strongly coupled with neutrinos. The LEP data on Z-decay are
 automatically reconciled with the presence of a strong enough
 \ne-majoron coupling, in the range $g_{ee} \sim 10^{-5} - 10^{-3}$,
 which leads to experimentally observable $2 \beta_{O \nu J}$ decay
 rates. Moreover, this could have consequences for \neu fluxes
 from gravitational collapse \cite{Berezhiani89}.
 %%We will refer to this region of g as the "strong" $\nu$-J coupling
 %%%%%%%%%%%%%%%%%%%%%%%%%%%%%%%%%%%%%%%%%%%%%%%%%%%%%%%%%%%%%%%
 The paper is organized as follows. In Sect.2 the status of neutrinoless
 double beta decays is reviewed in the context of the majoron models and
 the general idea on the possibility of strong $\nu$-J coupling is
 discussed.
 In Sect.3 different models of singlet (or dominantly singlet) majoron
 are presented as a realizations of this possibility. In Sect.4 we draw
 our conclusions.

 \section{General Considerations}

 In order to have strong neutrino-majoron $\nu_e$-J couplings,
 the scale of spontaneous lepton number violation has to be
 sufficiently low. This generates the majoron, denoted J,
 and originated from some neutral complex scalar singlet
 $\sigma$, defined through
 \beq
 \sigma=\frac{1}{\sqrt2}(v+\rho+iJ) \; .
 \eeq
 Below this scale $v$ the relevant terms in lagrangian are
 the neutrino mass term and the Yukawa couplings with the
 majoron J and its scalar partner $\rho$,
 \beq
 \label{1}
 L_{mass} = - \frac{m}{2} {\nu}^T_{e} \sigma_2 \nu_{e} \\ \quad
 L_{Yuk} = i \frac{g}{2} J {\nu}^T_{e} \sigma_2 \nu_{e}
 + \frac{g}{2}  \rho {\nu}_{e}^T \sigma_2 \nu_{e} %%\\\nonumber
 \eeq
 where $\sigma_2$ is the 2x2 Pauli matrix of charge conjugation
 and the majoron coupling g is given as
 \beq
 g = m / v
 \label{g}
 \eeq
 independently on the specifics of the majoron model
 \footnote{In the case of three neutrino flavours $m$
 is the $m_{ee}$ element of the neutrino Majorana mass
 matrix, and $g=g_{ee}$ is a corresponding coupling constant.}.

 Recent data on neutrinoless double beta decay
 %%from $^{76}Ge (2 \beta_{0\nu})$ and $^{32}Se (2 \beta_{0 \nu J})$
 yield, \cite{Klapdor91}
 $$
 m \lsim 1 - 3 eV \:\:\:\: \quad
 %%T_{1/2} > 1.2 \times 10^{24} yr
 \eqno{(3a)}
 $$
 $$
 g \lsim 10^{-4} - 10^{-3} \:\:\:\: \quad
 %%T_{1/2} > 1.6 \times 10^{21} yr
 \eqno{(3b)}
 $$
 where the bounds vary depending on the evaluation of nuclear matrix
 elements. The lower and upper values in (3a.b) correspond to the models
 \cite{vogelaa} and \cite{vogela} respectively. The sensitivity
 expected in planned experimental searches should improve these
 limits by an order of magnitude. Then \eq{g} implies that a
 positive result in $2 \beta_{0 \nu J}$ decay searches is
 expected only if $v \lsim 10 - 100$ KeV. This implies
 that in neutrinoless double beta decay not only the massless
 majoron J is emitted but also the accompanying $0^+$ scalar
 partner $\rho$ of mass $m_\rho \lsim v$. As a result we
 expect the existence of new features in the electron sum
 energy spectrum, as indicated in Fig. 1. Indeed, close to
 the endpoint Q, \ie between $Q-m_\rho$ and Q, only the massless
 majoron is emitted, while in the region from 0 to $Q-m_\rho$
 both $\rho$ and J are emitted. In principle this would imply
 a deficit of events close to the endpoint.

 It is interesting to note that
 for $v$ in this range the $2 \beta_{0 \nu J}$ decay should be
 faster as compared to $2 \beta_{0 \nu}$ one. Indeed, taking
 into account that the nuclear matrix elements for both processes
 are actually the same one has
 \beq
 \frac{\Gamma (2 \beta_{0 \nu J})}{\Gamma (2 \beta_{0 \nu})} \simeq
 \frac{1}{84 \pi^{2}} \left( \frac{Q}{v} \right)^{2} R (x)
 \label{5}
 \eeq
 where $R(x)$ is the corresponding phase space factor \cite{GGN}.
 Since the typical mass difference Q between the initial and final
 nuclei is a few Mev, it follows that the above ratio exceeds one for
 $v \lsim 10 - 100$ KeV.

 The simplest scheme containing a singlet majoron
 includes one isosinglet lepton which acquires a large mass from
 its interaction with $\sigma$, after spontaneous lepton
 number violation \cite{CMP}. If $v \gsim Q \sim 1$ MeV
 then the $2 \beta_{0 \nu J}$ decay is strongly suppressed
 due to the smallness of $g \sim \frac{m_\nu}{v} \lsim 10^{-5}
 - 10^{-6}$. On the other hand if the scale $v \ll Q$ the
 contributions of the mass eigenstate \neus {\sl cancel}
 so that both matrix elements $g_{ee}$ and $m_{ee}$ are
 zero and the probabilities of both types of neutrinoless
 $2 \beta$ decays, $2 \beta_{0 \nu J}$ and $2 \beta_{0 \nu}$
 vanish.
 In addition, this is in conflict with the cosmological
 nucleosynthesis constraints. Indeed, due to the strong $\nu J$
 interactions the right handed neutrinos together with majoron
 J and its scalar partner $\rho$ should come to equilibrium
 in the Early Universe before $ t=1 \:s$ and thereby significantly
 affect the stage of primordial nucleosynthesis. On the other hand
 the effective number of light neutrino species allowed by the
 primordial $^{4}$He abundance is $N_{\nu} < 3.4$ \cite{sch},
 in agreement with just the three conventional neutrinos.
 In contrast, in view of the astrophysical upper limit on
 the majoron coupling to electrons that follows from stellar
 energy loss arguments \cite{KIM}, the triplet and doublet majorons
 were considered as promising candidates for $2 \beta_{0 \nu J}$ decay.
 However, these models are now ruled out by LEP data.
 As a result we will consider alternative schemes with
 and without isosinglet \neus that can lead to substantial
 majoron emission rates in neutrinoless double beta decay
 (sect. 3).

 The lowest dimension effective operator, relevant for the
 generation of Majorana neutrino masses, is the following
 (see Fig. 2)
 \footnote{Generalization to the case of three lepton doublets
 is straightforward.}:
 \beq
 {\cal L} = G_J \ell^T \sigma_2 \tau_\alpha \ell
 H^T \tau_\alpha H \sigma + \rm{h.c.}
 \label{6}
 \eeq
 where $\sigma$ is an \21 singlet with L=-2,
 $\ell = (^{\nu_{e}}_{e})_{L}$ is a lepton doublet,
 $H=(^{H^{+}}_{H^{0}})$ is the standard Higgs doublet,
 and $\tau_{\alpha}$, $\alpha = 1,2,3$, are the isospin
 Pauli matrices. After \21 breaking by $\VEV{H^0} =
 (2 \sqrt{2} G_{F})^{-1/2} \simeq 188$ GeV, \eq{6}
 implies that $g = G_J \VEV{H}^2$ is the
 effective coupling constant g.

 As we emphasized above, the existence of strong \ne-J coupling
 requires a sufficiently low scale of lepton number violation
 $\VEV{\sigma} \ll \VEV{H}$. In this connection a question appears
 concerning the stability of this hierarchy. First note that this is
 not the usual the gauge hierarchy, as far as $\sigma$ is a gauge singlet.
 Its couplings with the Higgs field H in the Higgs potential can not be
 generated with gauge boson loops. On the other hand the radiative
 corrections to the mass term of $\sigma$ that follow from \eq{6}
 are of the order of $\delta M^{2}_{\sigma} \sim \frac{g^{2}}{32 \pi^{2}}
 \Lambda^2$, where $\Lambda$ is some physical cut off. Taking into
 account the experimental bounds on \neu mass and requiring
 that $v \sim \delta M_{\sigma}$ this corresponds to
 $g \sim 10^{-3} - 10^{-5}$ for $\Lambda$ in the range
 20 - 200 GeV. In particular models, considered in Sect. 3,
 there may be other contributions to $\delta M^{2}_{\sigma}$,
 that should be taken into account. Although we will not
 consider in great detail these radiative corrections, we would
 like stress that, from general arguments, in these singlet majoron
 models the possibility to have $\VEV{\sigma} \ll \VEV{H}$ does not
 necessarily require fine tuning, unlike the case of the triplet and
 doublet majoron models
 \footnote{For the latter a fine tuning, at least for the splitting
 of neutral component from the charged ones, was unavoidable.}.
 Although this question deserves the special consideration
 \cite{Berezhiani92}, here we wish to note that a cutoff in
 the required range may be naturally provided.

 We now consider mechanisms where the effective operator
 \eq{6} arises as the low energy limit of new renormalizable
 interactions, in analogy with the way in which the Fermi
 four fermion interaction emerges from W and Z boson exchange.

 \section{Models}

 The effective operator in \eq{6} may arise at the tree
 level from the exchange of a neutral heavy lepton (Fig. 3.a).
 Alternatively, it can appear as a radiative correction
 due to the effects of new Higgs bosons as in Fig. 3.b
 and Fig. 3.c. It may also arise from tree level heavy Higgs
 triplet exchange (Fig. 3.d). Let us discuss now these possibilities.

 \subsection{Singlet majoron in "$\mu$-Model"}

 This model \cite{CON} is a variant of the "seesaw"
 model with heavy Dirac lepton suggested in ref.
 \cite{SST1}. Here the operator \eq{6}
 can be effectively induced through the "exchange" of heavy
 Dirac neutral fermion carrying one unit of lepton number
 (see Fig. 3.a). The relevant terms in Lagrangian are the
 bare Dirac mass term and the Yukawa couplings
 \footnote{A possible $\nu^c \nu^c \sigma^*$ entry
 would not give a mass to the light \neus and can be
 forbidden by demanding supersymmetry.}
 \beq
 h_\nu \ell^T C \nu^c H + M \nu^c C S +
 f \sigma S^T C S + h.c.,
 \label{16}
 \eeq
 The first coupling generates the Dirac mass term $D = h_\nu \VEV{H}$,
 while the second term gives the Majorana mass term $\mu$
 for the isosinglet $S$ that violates lepton number by
 two units, $\mu = f \VEV{\sigma}$. The full mass matrix
 in Majorana basis $\nu, \nu^c, S$ can be written as
 \footnote{For the generalization to three neutrino
 families one introduces six isosinglet neutrinos,
 three $\nu^c$-type and three $S$-type.}
 \beq
 \left( \begin{array}{ccc}
 0 & D & 0\\
 D & 0 & M\\
 0 & M & \mu \\
 \end{array} \right)
 \label{17}
 \eeq
 For small values of the parameter $\mu \ll D \ll M$,
 the heavy leptons here are of Quasi-Dirac type and
 the Majorana mass of the light, mostly isodoublet
 \neu $\nu_{L}$ arising from \eq{17} and the \ne-
 majoron coupling constant are, respectively
 \beq
 m = \mu \left( \frac{D}{M} \right)^{2} , \quad g = \frac{m}{v} =
 f \left( \frac{D}{M} \right)^{2}
 \label{18}
 \eeq
 Note the different relationship between $m_\nu$
 and $\mu \lsim \VEV{\sigma}$, the lepton number
 breaking scale. This is a crucial feature of this
 model, which contrasts with the simplest seesaw model.
 It follows from the fact that the model contains a
 quasi-Dirac heavy lepton whose mass $\sim M$, is invariant
 under lepton number and unrelated to \neu masses. In contrast,
 the minimal seesaw model has a heavy Majorana lepton whose
 mass is inversely related to that of the isodoublet neutrino.
 In both cases the heavy
 lepton admixture in the weak charged current
 is determined by the ratio $D/M$. However, in the
 "$\mu$-model" this value is restricted only by weak
 universality constraints, and not by limits on
 \neu masses \cite{fae}
 \footnote{The crucial point is that the heavy
 lepton mass here arises mostly from the entry
 M, invariant under lepton number, unlike the
 case of the seesaw model. As a result M can be
 relatively low without implying too large $m_\nu$
 values. In fact, in the limit where lepton number is
 exact \neu masses are strictly forbidden \cite{SST1}.}
 As a result a \ne-J coupling as large as $10^{-3}$
 is can be achieved for values of $D/M \lsim 0.1$,
 well consistent with present weak universality
 constraints \cite{fae}.

 The phenomenology of this model was considered
 in detail in several papers \cite{fae}. For masses
 below $m_Z$ the heavy quasi-Dirac leptons may be
 singly produced at LEP1, giving rise to striking
 events characterized by a large amount of missing energy
 \cite{CERN1,CERN2}. The existence of such neutral heavy
 leptons at higher masses can at present only be probed
 due to their indirect effects. For example, if we
 include mixing between the various generations
 we have the interesting possibility of flavour
 and/or CP violation \cite{BER,CP,CPa} even in the
 limit where $\mu \ra 0$ and the isodoublet \neus
 become strictly massless. As a result, processes
 \sa $\mu \ra e \gamma$, $\tau \ra e \gamma$,
 $\mu -e$ conversion in nuclei, $Z \ra e \tau$,
 \etc are all allowed. Moreover, the corresponding
 rates are restricted only by the precision of
 weak universality tests. As a result, they all
 can be within the sensitivity of present
 experiments as well as those expected at
 the upcoming tau factory \cite{3E}.

 \subsection{Singlet Majoron in Zee Model}

 This model is a variant to the original Zee model \cite{zee}
 which introduces the spontaneous violation of lepton number
 \cite{BarbieriHall} and, as a consequence, the existence of
 a massless pseudoscalar Majoron. The relevant
 terms in Lagrangian are the Yukawa couplings
 \footnote{For simplicity we assume that
 all of the Yukawa coupling constants are real.}
 \beq
 \frac{m_i}{\VEV{H}}{\bar{\ell}}_i H e_{Ri} +
 h_{ij}{\bar{\ell}}_i \phi e_{Ri} +
 f_{ij} \ell_i^T C i \tau_2 \ell_j \eta^+ + h.c.
 \label{yuk}
 \eeq
 where $i,j= e,\mu,\tau$. The first term is the canonical
 one responsible for generating the charged lepton masses
 $m_i$ when the $SU(2) \ot U(1)_Y$  \gau symmetry is broken by
 $\VEV{H}$. The additional couplings involve
 another Higgs doublet $\phi$ as well as the Zee singlet and are
 specified by matrices $f,g$ (in generation space) $f$ being
 antisymmetric. In addition we use the following quartic term
 in the scalar potential
 \beq
 \lambda H \sigma_2 \phi \eta^{+} \sigma \: + h.c.
 \eeq
 instead of the usual cubic term $\Lambda H \sigma_2 \phi \eta^{+}$
 that would explicitly violate lepton number. The quartic coupling
 fixes $L(\sigma)=2=-L(\eta)$ and induces a mixing between the
 physical singly charged scalars which plays a crucial role in
 the radiative generation of \neu mass and of the
 \ne-J coupling. Indeed, the effective operator in
 \eq{6} arises at one loop level, See Fig. 3.b
 and has the following associated coupling constant
 \beq
 g = G_{J} \VEV{H}^2 = \frac{h_{ei} f_{ie} \lambda}{16 \pi^{2}}
 \frac{\VEV{H} m_i } {M^{2}_{\phi} - M^{2}_{\eta}}
 \ln \frac{M^{2}_{\phi}}{M^{2}_{\eta}}
 \label{G}
 \eeq
 where $M^{2}_{\phi}$ and $M^{2}_{\eta}$ are the eigenvalues of
 the mass matrix of charged scalars $\phi^{-}$ and $\eta^{+}$
 in the limit of small mixing. From Fig. 3.b it is clear that
 if we require natural flavour conservation in neutral Higgs
 couplings ($h_{ij}$ diagonal) then $g = 0$. Therefore we now
 relax this requirement and allow nondiagonal flavour
 changing couplings for $\phi$ in \eq{G}.

 In order to estimate the maximal possible value of
 the induced \ne-J coupling constant in this model
 we first note that the dominant contribution will
 come from $\tau$ lepton exchange in the loop.
 Next, we must discuss the relevant constraints. First
 note that the exchange of the scalar $\eta^{+}$
 gives a contribution to the decay width of the
 $\tau$-lepton, $\tau \ra e \bar{\nu}_{e} \nu_{\tau}$
 given as $\Gamma_{\tau} = \Gamma^{SM}_{\tau} (1 + 2 \rho_{\tau})$,
 where $\rho_{\tau} = \frac{f^{2}}{M^{2}} / 8 \sqrt{2} G_{F}$. The
 present experimental limit $\rho_{\tau} \lsim 2.5 \%$ then implies
 $f/M_{\eta}^{+} \lsim \frac{0.33 }{\VEV{H}}$. Similarly,
 $\phi^{+}$ exchange leads to a new contribution to the process
 $e^{+} e^{-} \ra \nu_{\tau} \bar{\nu}_{\tau}$. While this
 process is invisible, the related process $e^{+} e^{-} \ra
 \nu_{\tau} \bar{\nu}_{\tau} \gamma$ can be seen. The limit
 from the ASP experiment \cite{Hearty87} then implies
 $\frac{8}{M_{\phi^{+}}} \lsim \frac{ 2.2}{\VEV{H}}$.
 One can now easily estimate that the maximum
 attainable values for g (which occurs for
 $M_{\phi} \simeq M_{\eta}$) consistent with the
 restrictions above and with the recent LEP limits
 on charged scalar masses is
 \beq
 g = G_{J} \VEV{H}^2 \lsim \frac{f h \lambda}{16 \pi^{2}}
 \frac{m_\tau}{M_{\eta}} \frac{\VEV{H} }{M_{\phi}} \lsim
 4 \times 10^{-5} \lambda
 \label{GG}
 \eeq
 So, in this model, provided that $\lambda < 1$, a
 sufficiently large value for the \ne-J coupling constant
 within the sensitivity of neutrinoless double beta decay
 experiments can only be obtained by adjusting all the
 parameters at their allowed borders.

 \subsection{Singlet Majoron in a Coloured Zee Model}

 This is a variant of the previous model which
 can easily allow for larger \ne-J coupling in
 $2 \beta_{0 \nu J}$ decay. The relevant terms
 in the Lagrangian are
 \beq
 h_{ij} {\ell}_i^T C b^c_{Lij} \phi_D +
 f_{ij}{\ell}_i^T C Q_{Lj} \eta_S +
 \lambda H \sigma_2 \phi_D \eta_S \sigma \: + h.c.
 \label{cz}
 \eeq
 where i,j= 1,2,3 and $\phi_D = (^{\phi^{2/3}}_{\phi^{-1/3}})$
 and $\eta_S$ are colour triplet leptoquark scalars:
 $\phi_D$ is an \21 doublet
 with Y=1/3 and L=-1 while $\eta_S$ is a singlet with Y=2/3 and
 also L=-1. Again lepton number is spontaneously broken
 by $\VEV{\sigma}$ and this generates the majoron, as
 in the previous case. The key contribution to the effective
 operator in \eq{6} now arises (See Fig. 3.c) from
 b-quark exchange in the loop. One obtains:
 \beq
 g = G_{J} \VEV{H}^2 \approx \frac{3 h_{eb} f_{be} \lambda}{16 \pi^{2}}
 \frac{m_b \VEV{H}} {M^{2}_{\phi_D} - M^{2}_{\eta_S}}
 \ln \frac{M^{2}_{\phi_D}}{M^{2}_{\eta_S}}
 \label{GC}
 \eeq
 in a similar notation as above. This contribution
 is enhanced with respect to the one in \eq{G} by a
 factor $\sim 3 m_b/m_\tau$, 3 coming from colour.
 In addition, flavour changing neutral current mediated
 interactions are automatically forbidden at the tree
 level since there is only one isodoublet Higgs contributing
 to generating the masses of the charged fermions.
 The constraints on the allowed \ne- J coupling strength
 now arise from the contribution of $\eta^{+ 1/3}$ exchange in
 b-meson semileptonic decays. Considering that the uncertainties
 in the limits on b-meson decays are somewhat higher than in the
 $\tau$ lepton case one can easily obtain the maximum attainable
 values for the \ne- J coupling consistent with experiment as
 \beq
 g = G_{J} \VEV{H}^2 \lsim 10^{-3} \lambda
 \label{GGG}
 \eeq
 Thus, for reasonable choices of parameters, we
 can have in this model a large enough value for the
 \ne- J coupling constant as to be of interest in
 neutrinoless double beta decay experiments.

 \subsection{Singlet-Triplet Majoron }

 For completeness we note that it is also possible
 to generate the desired $\nu$-J interaction in a
 scheme where the majoron is not a pure \21 singlet.
 The most straightforward example is to invoke a
 triplet-singlet scheme, suggested in \cite{774}.
 The model contains the following couplings
 \beq
 f \ell^T \sigma_2 \tau_{\alpha} \Phi_{\alpha} \ell +
 \lambda \Phi^{\ast}_{\alpha} H \tau_{\alpha} H \sigma + h.c.
 \label{8}
 \eeq
 where $\Phi_{a} = (\Phi^{++}, \Phi^{+}, \Phi^{0}), a=1,2,3$
 is a weak isospin triplet carrying hypercharge Y = 2 and
 lepton number L=-2. In this case the $\nu$-J coupling
 arises from $\Phi$ exchange, as shown in Fig. 3.d.
 In order to avoid an excessive contribution to the
 invisible Z-boson width, the neutral component $\Phi^0$
 should be heavier than about 40 GeV, whereas the lower limits
 on the charged components should be somewhat stronger.

 In such a model the majoron is given as \cite{774}
 \beq
 J \simeq \frac{1}{v}
 (\VEV{\sigma} \: Im \sigma + \VEV{\Phi} \: Im \Phi^0)
 \eeq
 where $v=\sqrt{\VEV{\sigma}^{2} + \VEV{\Phi}^{2}}$ is the scale of
 lepton number violation. In order to avoid an excessive contribution
 to the invisible Z width we need to require $\VEV{\Phi} < \VEV{\sigma}
 < 10 - 100$ keV. The first inequality may be achieved if the VEV
 of the triplet arises only from the quartic interaction term
 present in \eq{8}. Such induced VEV is given as
 $\VEV{\Phi} \sim \frac{\lambda v \VEV{H}^2}{M_{\Phi}^2}$.
 The second inequality is needed in order to have a sizeable
 \ne \ne J coupling. With this it follows that the
 electron-majoron coupling is automatically too small
 to give any significant contribution to stellar energy
 loss. Although phenomenologically viable, we do not favour
 this model because of the fine-tunings necessary in order to
 have a sizeable Majoron coupling, a situation reminiscent
 of the old triplet Majoron model.

 \section{Conclusions}

 LEP data on Z-decays do not exclude the possibility of
 observable rates for majoron emission in neutrinoless
 double beta decays. The majoron coupling to the Z boson
 is determined by a \gau coupling and weak isospin, whereas
 its coupling to neutrinos responsible for majoron
 double beta decays is of Yukawa origin. In general
 these couplings are not related. We have presented
 models where the LEP restrictions are avoided because
 the majoron is a singlet, or dominantly singlet under
 the electroweak \21 gauge symmetry. At the same time
 the effective coupling g of the singlet majoron to the
 \ne can be as large as $g \sim 10^{-3} - 10^{-5}$.
 The key points of these models with large neutrino-
 majoron couplings are the existence of new heavy
 leptons and/or scalar bosons at the electroweak
 scale with the spontaneous violation of lepton
 number occurring below 10 - 100 keV. In the most
 appealing model (section 3.1) the effective
 neutrino-majoron coupling arises from the exchange
 of a neutral Quasi-Dirac heavy lepton (Fig 3.a).
 We also discussed the possibility of inducing this
 coupling from radiative corrections associated with
 additional Higgs bosons (Fig 3.b and 3.c) or tree-level
 Higgs exchange (Fig. 3.d). Of these models where the
 enhanced majoron-neutrino coupling arises mostly
 from scalar boson exchange the best possibility
 is the coloured Zee model.

 We thank A S Barabash, G Fiorentini, M I Vysotsky and
 M G Shchepkin for discussions.

 %%%%%%%%%%%%%%%%%%%%%%%%%%%%%%%%%%%%
 \newpage
 \bibliographystyle{ansrt}
 \bibliography{biblio}
 \end{document}